\documentclass[letterpaper,twocolumn,10pt]{article}
\usepackage{usenix,epsfig}

\usepackage{graphicx}

\graphicspath{{images/}}
\usepackage{url}
\usepackage{color}
\usepackage{listings}
\usepackage{tabularx}
\usepackage{ragged2e}
\usepackage{xcolor,colortbl}
\usepackage{multirow}
\usepackage{subfigure}
\usepackage[noadjust]{cite}
\usepackage{algorithm}
\usepackage[noend]{algpseudocode}

\usepackage[utf8]{inputenc}
\usepackage{cleveref}
\crefname{section}{§}{§§}
\Crefname{section}{§}{§§}
\usepackage[shortcuts]{extdash}
\usepackage{balance}

\usepackage[utf8]{inputenc}
\usepackage{listings}
\usepackage{fancybox}
\usepackage{fancyhdr}
\usepackage{fancyvrb}

\usepackage{etoolbox}
\newtoggle{our_comments}
\toggletrue{our_comments}

\def\BState{\State\hskip-\ALG@thistlm}

\definecolor{Gray}{gray}{0.85}

\newcolumntype{Y}{>{\RaggedRight\arraybackslash}X}

\usepackage{setspace}
\definecolor{dkgreen}{rgb}{0,0.6,0}
\definecolor{gray}{rgb}{0.5,0.5,0.5}
\definecolor{mauve}{rgb}{0.58,0,0.82}

\usepackage{enumitem}
\setlist{leftmargin=4ex, topsep=0.5ex,noitemsep}

\begin{document}

\title{Unified Management and Optimization of Edge-Cloud IoT Applications \vspace*{-2.5ex}}
\author{
	{\rm Shadi A. Noghabi}\\
	UIUC \\
	abdolla2@illinois.edu
	\and
	{\rm Jack Kolb}\\
	Berkeley\\
	jkolb@cs.berkeley.edu
	\and
	{\rm Peter Bodik, Eduardo Cuervo}\\
	Microsoft Research\\
	\{peterb,cuervo\}@microsoft.com 	
} 

\maketitle

\section{Introduction}
\label{sec:intro}
Internet of Things (IoT) solutions are growing more popular and disruptively in scale. Gartner Inc. estimates that the number of deployed IoT devices will grow from 5 Billion in 2015 to 25 Billion in 2020 creating a multi-billion dollar market \cite{iotScale, iotScale2}.
This growth has led to many industrial services offering to simplify development and management of IoT applications, e.g., C3 IoT and IoT frameworks from all major cloud providers \cite{azure-iot, google-iot, aws-iot, c3}.
Initially, AWS, Azure and Google offered managed solutions based completely in the cloud.
An IoT management service (such as IoT Hub from Azure) securely manages the remote devices, their configuration, and receives the sensor data. The rest of the processing pipeline is built using regular cloud services for computation (Azure Function, or Stream Analytics), communication (Event Hub), and storage (CosmosDB, Blobs).

As the IoT solutions scale up in the number of devices and messages sent, there is greater need to process data closer to the IoT sensors using the edge. Processing on the edge has several advantages including: reducing end-to-end latency especially when the application is actually controlling the devices (e.g., shutting down operations in case of failures), providing continuation of service despite low connectivity to the cloud, reducing the bandwidth usage, and reducing monetary cost of using cloud services \cite{to-edge, edge-vision, cloudlet-bahl, fog-iot}. 


However, the IoT industry is still in its infancy. Building an IoT solution comes with a lot of challenges and complexities in deployment, monitoring, and optimizing the end-to-end solution.
Even a simple remote monitoring application will consist of several components on the edge to receive, preprocess, and send the data to the cloud along with services in the cloud for additional processing, machine learning, publish-subscribe systems, and storage. Deploying such an application requires a lot of error-prone and complicated scripting. Since each cloud service has grown organically and independently, there is a great diversity among them in terms of their cost models, monitoring, API, etc, and this diversity has created many compatibility constraints across services.

Moreover, naive implementations of IoT applications can often be very inefficient both in terms of cost and performance. For example, processing all in the cloud a) requires a lot of bandwidth to send sensor measurements (typically encoded as JSON strings), and b) becomes expensive for large datasets.
Therefore, developers manually optimize the processing pipeline by batching data for upload, compressing data, and determining component placements. Making these decisions is non-trivial because the best configuration depends on the available resources, the workload, and accurate estimations of a new configuration.
Also, companies deploy a wide range of heterogeneous edge devices~\cite{AzureDevices} (from Raspberry Pi to large servers) and use different connectivity options (optical, WiFi, cellular, or satellite) further complicating this decision.

We present a system, Steel, that abstracts the complexities of building and optimizing IoT applications while maintaining the flexibility of using cloud services and the edge.
Steel allows users to declaratively describe the IoT application. Then, Steel manages the deployment, monitoring, and automatic cost and performance optimizations in response to changes in workload, resource demands, or failures. In this work, we focus on optimizing \textit{placement}, i.e, where to place each component, and \textit{communication}, i.e., how to optimize the edge-cloud link.

\begin{figure}[t]
	\centering
	\includegraphics[width=0.9\linewidth]{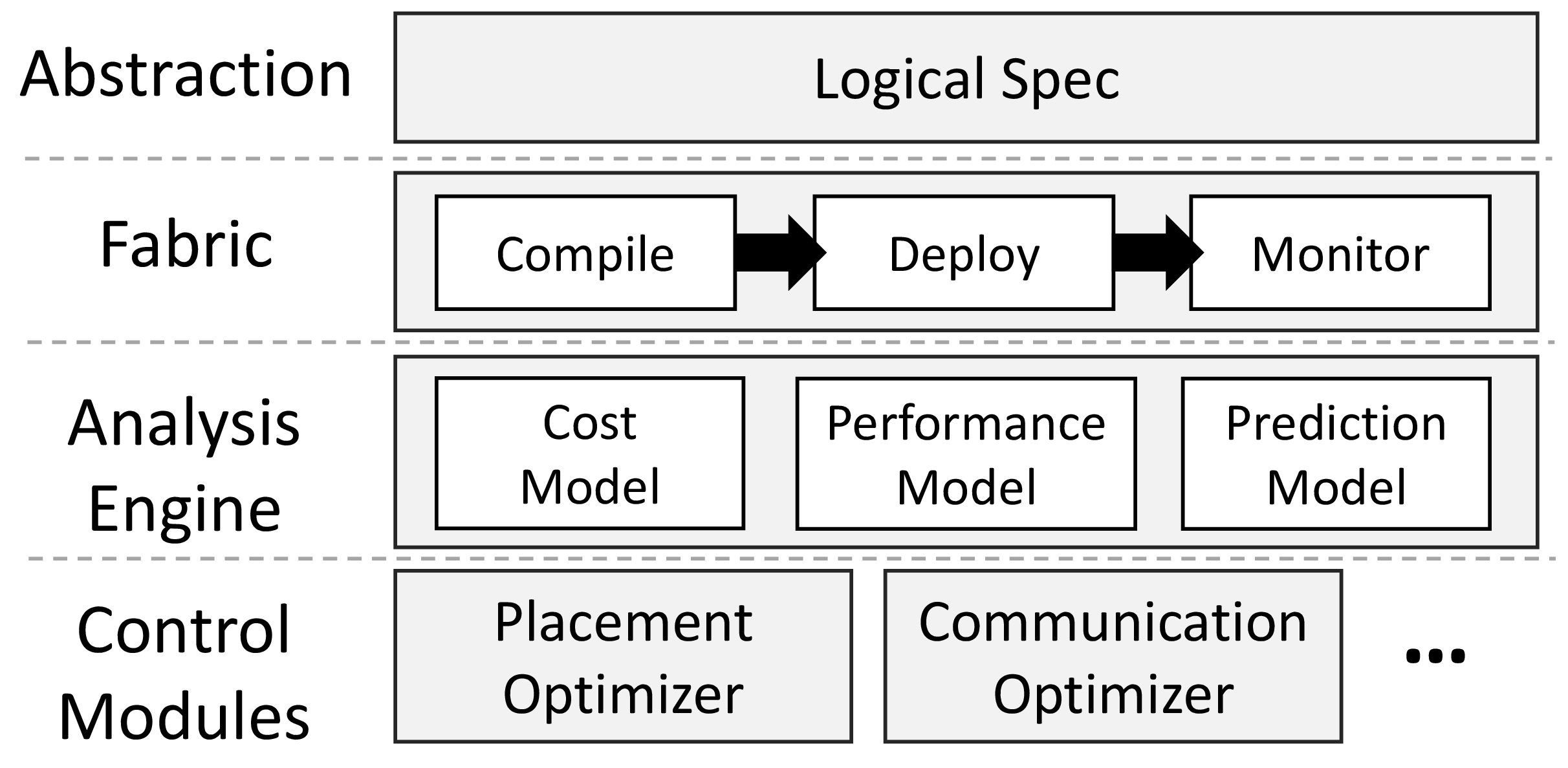}
	\vspace*{-2ex}
	\caption{High level architecture of Steel.}
	\label{fig:architecture}
	\vspace*{-2ex}
\end{figure}

We implemented Steel on top of an IoT edge/cloud stack from one of the leading cloud providers (Azure IoT stack) and demonstrate that we can deploy realistic IoT applications from a simple job spec. 
The evaluation of our placement and communication optimizations based on real-world workloads from two production IoT deployments shows that they effectively adapt to dramatic changes in workload and resource capacities. Our placement optimization accurately predicts cost for new configurations and iteratively reduces cost up to 40\% at each iteration, while also increasing the resource utilization of the edge devices up to 75\%. Our adaptive scheme reacts to changes in available bandwidth within 1 second. Even in presence of a 20x reduction in bandwidth cap, the data sent is adapted to stay within the limit, with a peek of 2.3x extra bandwidth usage.


\section{Design of Steel}

As shown in Figure \ref{fig:architecture}, Steel consists of four main layers: abstraction, Fabric, Analysis Engine, and Control Modules, described in more detail below.

\noindent\textbf{Abstraction:} Instead of directly deploying individual modules, developers describe their application in a \emph{logical spec}. The spec defines the data  source for the application (e.g., all temperature sensors in building 43), the processing components (e.g., computing average temperature), and how the components are connected into a directed acyclic graph. The spec abstracts the data flow (connections between components) while providing full flexibility on the internal computation in each component, supporting both user-defined code and cloud services.

\noindent\textbf{Fabric:} The IoT Fabric materializes the logical spec into an actual physical deployment. The Fabric takes the logical spec as an input; \textit{compiles} the spec into a physical layout while adding other necessary components (such as data compression and decompression); \textit{deploys} the application automatically and in parallel across the entire edge-cloud environment; and then \textit{monitors} the application end-to-end reporting performance and resource usage metrics. 

\noindent\textbf{Analysis Engine:} 
Even though most IoT applications have a common and simple pipeline (measure, process, store, and visualize), analyzing them is non-trivial due to the inherent complexities of the environment. First, IoT applications typically use over four distinct cloud services \cite{AzureDevices}, each with diverse \textit{pricing structures}: different metrics for charge (bytes, CPU, IO), complex and opaque metrics (SU for Stream Analytics), various granularities, provisioned resources, etc.  
Secondly, due to the limited yet heterogeneous resources on the edge, from raspberry pi to large servers, it is difficult to predict the resource demand and performance of components on a new device.
To hide these complexities, the Analysis engine builds both cost and performance models, by using monitoring metrics and proactive shadowing (creating a shadow component in a new location). Using the models, it creates a prediction engine to answer \textit{"what-if"} questions, predicting cost and performance of changes without applying the change.

\noindent\textbf{Control Modules:} On top of the fabric, we run several \emph{Control Modules} that monitor the application and adjust its configuration at runtime in response to changes in workload and environment. The goal of these modules is to remove the burden of common manual optimizations from developers. In this work, we focus on two most popular optimizations: \textit{placement} and \textit{communication}.

\subsection{Placement Optimization}

\begin{figure}[t]
	\centering
	\vspace*{-4ex}
	\includegraphics[width=0.9\linewidth]{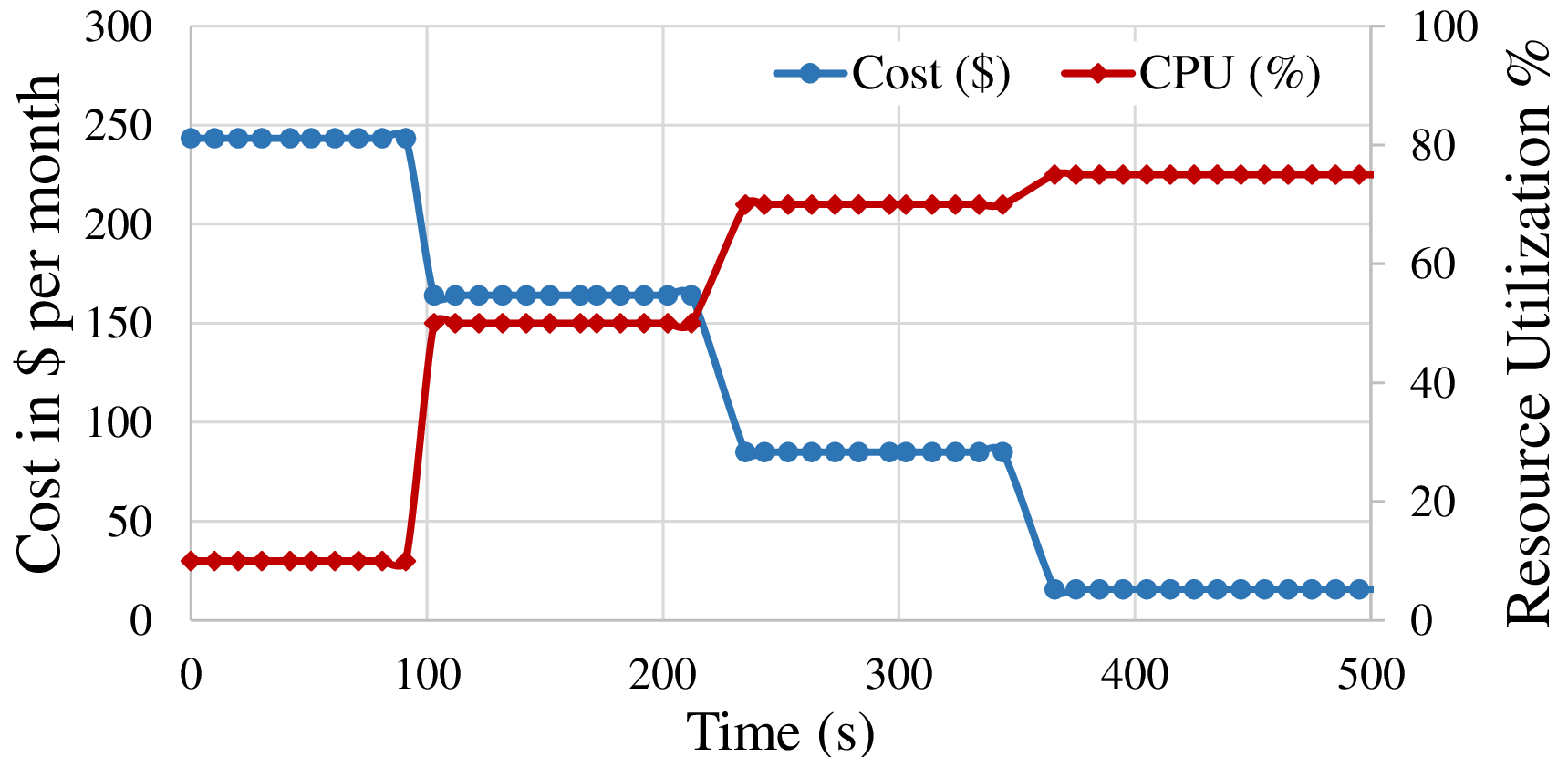}
	\vspace*{-3ex}
	\caption{Reduction of price and improvement of edge utilization with the placement optimizer over time.}
	\vspace*{-2ex}
	
	\label{fig:price-time-res}
\end{figure}

One of the common and complex optimizations for IoT applications is the placement of different components. In a typical industrial setting with hundreds of thousands of sensors and hundreds of edges, manually optimizing the placement, if even possible, is sub-optimal, error-prone and not adaptive. We developed a placement optimizer that takes the cost and performance models of various components and automatically optimizes the placement. 

The optimizer starts all in the cloud. Then, in a greedy fashion, it finds the most expensive components and suggests new placements for each. Using the analysis engine, it predicts the impact of a move and moves the least cost-efficient ones. This loop continues until there is no more capacity on the edges for making moves. Figure \ref{fig:price-time-res} shows a sample run of a multi-stage application. At each iteration, the overall cost is reduced  by $~ 40\%$ as components move from the cloud to the edge. Along with that, the resource utilization of the edge increases from $10\%$ to $75\%$.

\subsection{Communication Optimization}

\begin{figure}[t]
	\centering
	\vspace*{-4ex}
	\includegraphics[width=0.9\linewidth]{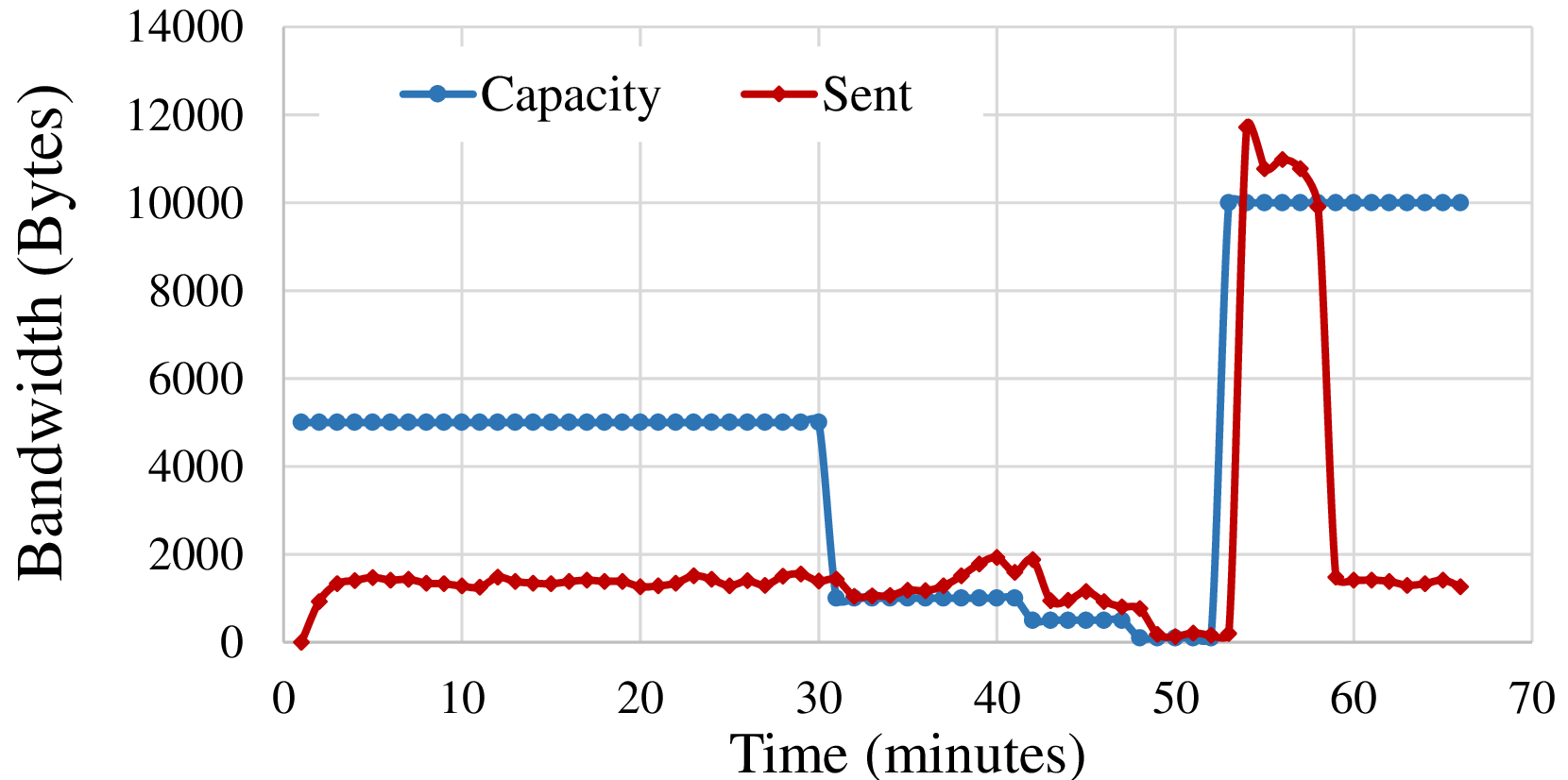}
	\vspace*{-3ex}
	\caption{Bandwidth consumed by the communications optimizer as network constraints are changed}
	\vspace*{-2ex}
	\label{fig:compressionExperiment}
\end{figure}


IoT deployments are often characterized by limited bandwidth or intermittent connectivity to the larger Internet, making the network a critical factor for both performance and cost. This optimizer hides the complexities of communication by transparently using an appropriate strategy for transforming and encoding data batches. 

In this work, we specifically consider the case of timeseries data emitted by sensors, arguably the dominant form of traffic that flows from the edge to the cloud in the vast majority of IoT deployments today. The Communication optimizer intelligently and adaptively chooses the best batching and compression mechanism, from a large search space of potential choices (e.g, binary packing, gzip, delta encoding, etc.). This is done on the fly, as network conditions change, to cope with variability or outages in network service, with the goal of minimizing cost.

Figure \ref{fig:compressionExperiment} shows the compression control module in action with replayed messages drawn from a real-world smart factory deployment. This trace begins with a generous bandwidth cap (shown in red) that is later lowered, mimicking a temporary congestion. The optimizer adapts the encoding scheme and message batch size to limit its consumption of bandwidth (shown in blue). Although bandwidth consumption does not stay strictly under the specified cap, it does stay reasonably close (up to 2.3x at peek). When cap is finally increased, the optimizer drains its buffers of backlogged data, which explains the brief peak in bandwidth consumption before it levels off to a steady rate at the end of the trace.

\bibliographystyle{acm}
\def\bibfont{\footnotesize}

\vspace*{-2ex}	
\bibliography{references}

\balance

\end{document}